\documentclass[11pt,twoside]{article}
\usepackage{APN5proc}

\resetcounters

\begin{document}

\title{From Bipolar to Elliptical: Morphological Changes
in the Temporal Evolution of PN}
\author{M. Huarte Espinosa$^{1,2}$, A.~Frank$^1$, B.~Balick$^3$,
O.~De~Marco$^4$, J.~H. Kastner$^5$, R.~Sahai$^6$ and
E.~G.~Blackman$^1$
\affil{$^1$Department of Physics and Astronomy, 
University of Rochester, 600 Wilson Boulevard,
Rochester, NY, 14627-0171}
\affil{$^2$Kavli Institute for Cosmology Cambridge, 
Madingley Road, Cambridge CB3 0HA, UK} 
\affil{$^3$Department of Astronomy, University of Washington, Seattle, WA 98195}
\affil{$^4$Department of Physics, Macquaurie University, 
Sydney NSW 2109, Australia}
\affil{$^5$Rochester Institute of Technology, 54 Lomb Memorial Drive, Rochester, 
NY 14623, USA}
\affil{$^6$NASA/JPL, 4800 Oak Grove Drive, Pasadena, CA 1109, USA}
}

\begin{abstract}
Proto-planetary nebulae (pPN) and planetary nebulae (PN) seem to be
formed by interacting winds from asymptotic giant branch (AGB)
stars.  The observational issue that most pPN are bipolar but most
older PN are elliptical is addressed. We present 2.5D~hydrodynamical
numerical simulations of episodic cooling interacting winds to
investigate the long term evolution of PN morphologies. We track
wind acceleration, decrease in mass-loss and episodic change in
wind geometry from spherical (AGB) to collimated (pPN) and back to
spherical again (PN). This outflow sequence is found to produce
realistic PN dynamics and morphological histories. Effects from
different AGB distributions and jet duty cycles are also investigated.\\
\noindent{\bf Keywords.}\hspace{10pt}Planetary Nebulae
\end{abstract}

\section{Introduction}
Planetary Nebulae are thick ionized plasma clouds that expand at
$\sim\,$20\,km\,s$^{-1}$ away from an old, hot, intermediate-mass
star. The nebulae show bipolar, elliptical, point symmetric,
irregular, spherical and quadrupolar morphologies \citep[for a
review see][]{balick02}. The interacting stellar wind model
\citep*[ISW;~][]{kwok78} suggest that spherical PN form by the
collision of the dust slow dense shell around an AGB star and the
tenuous fast wind that it blows at the post-AGB phase.  As opposed
to PN, AGB envelopes are typically spherical, therefore, PN must be 
shaped as they evolve, by some mechanisms which are not
clear yet \citep*{balick02}.
Generalized ISW models propose that bipolar PN form by the interaction
of the post-AGB fast wind and either a toroidal AGN envelope
\citep*[e.g.][]{frank94} or an aspherical AGB wind
\citep*[e.g.][]{icke92}. A binary system may cause asymmetries
in the AGB wind; where either an AGB interacts with a companion or
an AGB and its companion share a common envelope evolution \citep*[for
a review see][]{demarco09}.  Magnetic fields \citep*[e.g.][]{blackman01},
the rotation of the AGB \citep*[e.g.][]{ggs97} and photoionization
heating from the central star \citep*[e.g.][]{mellema97} have also
been considered in PN shaping.

Jets are evident in high resolution sensitive observations of many
pPN and young PN \citep*[e.g.][]{balick00,sahai00}.  The outflows
appear to be bipolar, collimated and launched at $\sim\,$200\,km\,s$^{-1}$
from the vicinities of the central star.  Jets are thought to shape
PN, to form knots in the nebulae and also to yield point symmetric
objects \citep*[e.g.][]{sahai98b}.

Here we present numerical simulations of episodic interacting winds
to address the observational issue that more than 50\,\% of pPN are
bipolar but more than 50\% of older PN are elliptical.

\section{Model and methodology}

Numerical simulations of interacting stellar winds are presented.
We track wind acceleration, mass-loss history and
episodic change in wind geometry.
The equations of radiative hydrodynamics are solved in two-dimensions,
with axisymmetry conditions, using the adaptive mesh
refinement (AMR) code AstroBEAR 
\citep{bear}. We use the tables of \citet*{dm} to simulate optically thin
cooling, ionization of H and He, and H$_{2}$ chemistry too.  No gravitational
or viscous or magnetic processes are considered.

The computational domain is a square representing 1\,pc$^2$. We
use extrapolation boundary conditions in the upper, the lower and
the right domain edges, and reflective conditions in the left edge.
Cylindrical coordinates are used with the origin at the middle of
the left boundary, \hbox{$r \in ($0,$\sqrt{2})$\,pc} and \hbox{$\theta
\in (-\pi/$2,~$\pi/$2$)$\,rad}. The grid has 128$^2$\,coarse cells
and two AMR levels; an effective resolution of $\sim\,$400\,AU.
We use BlueHive\footnote{
https://www.rochester.edu/its/web/wiki/crc/index.php/BlueHive\_Cluster},
an IBM parallel cluster of the Center for Research Computing of the University of Rochester,
to run each simulation for about 20\,hrs, using 16~processors.

\subsection{Wind episodes}

We consider three wind episodes: the isotropic AGB wind, the
collimated jet and the isotropic fast (post-AGB) wind.  The AGB
is the initial condition. Simulation~1 follows the
interaction of the AGB wind and the jet, whereas Simulation~2 follows
the interaction of the AGB and the fast wind.  Simulation~3
tracks
the interaction of the AGB wind, the jet which is ejected afterwards
and the fast wind which comes after the jet (see Table~1).

The AGB wind is set throughout the domain with an
ideal gas equation of state \hbox{($\gamma=\,$5$/$3)}, a
temperature of 500\,K, a radial velocity of 10\,km\,s$^{-1}$ and a
mass-loss of 10$^{-5}$\,M$_{\odot}$\,yr$^{-1}$. The
jet is injected for 108\,yr, only, in cells where $r<\,\,$6000\,AU,
with a collimated horizontal velocity of 200\,km\,s$^{-1}$, the
AGB's temperature and half of the AGB's density.  The isotropic
fast wind is continuously injected at $r<\,\,$6000\,AU, with a mass-loss
that decreases in time from 5$\times$10$^{-7}$~to
\hbox{5$\times$10$^{-9}$\,M$_{\odot}$\,yr$^{-1}$}, following the
model of \citet{perinotto}. The fast wind accelerates from
200~to 2000\,km\,s$^{-1}$, maintaining a constant ram
pressure, and we keep a Mach~20 condition in the injection region.

Additionally, Simulations~4, 5,~and~6 follow the interaction between
the jet, the fast wind or both, respectively (see Table~1), and an
initial AGB with a pole-to-equator density contrast of
1$/$2. We use the toroidal density distribution in equations~(1)
and~(2) of \citet{frank94}\footnote{
We use the referred equations for $\alpha=\,$1$/$2 and $\beta=\,$1.}.
Finally, in Simulation~7, the fast wind interacts with a spherical AGB wind
having a pole-to-equator velocity contrast of~2. This is modeled by multiplying
the AGB's radial velocity by $1+e^{-[\tan^{-1}(|y/x|)/0.3]^2}$
(see Table~1).

\vskip-0.5cm
\begin{table}[ht]
   \caption{Simulations and parameters.}
   \smallskip
   \begin{center}
      {\small
         \begin{tabular}{clcc}
            \tableline 
            \noalign{\smallskip}
            Simulation    &AGB wind   &Jet duration                 &Fast wind duration  \\
            ~~~~          &~~~~~form   &$[\times$108\,yr$]$ &$[\times$1000\,yr$]$   \\
            \noalign{\smallskip}
            \tableline 
            \noalign{\smallskip}
            1          &spherical              &1                &~\,0.0  \\
            2          &spherical              &0                &13.0    \\
            3          &spherical              &1                &10.7    \\
            4          &toroidal~$\rho$        &1                &~\,0.0  \\
            5          &toroidal~$\rho$        &0                &~\,3.8  \\
            6          &toroidal~$\rho$        &1                &~\,6.0  \\
            7          &aspherical~${\bf v}$   &0                &13.0    \\
            \noalign{\smallskip} 
            \tableline 
         \end{tabular} 
      } 
   \end{center} 
\end{table}

\section{Results and discussion}
\vskip-0.25cm
We present a summary of the simulation results, for  details see
Huarte~Espinosa et~al. 2010 (in prep.).  Figure~1, top row,
shows the evolution of Simulation~1.  The jet collides with the AGB
envelope, drives a bow~shock and forms a central elliptical cavity.
Jet injection ceases at 108\,yr and gas expands passively afterwards.
The lobe develops a bipolar morphology with a monotonically increasing
aspect ratio (i.e. the ratio of its longer dimension to its shorter
dimension) that reaches 4.5 in 13283\,yr. Conversely, Simulation~2
(middle row) follows the ISW model \citep*[][]{kwok78} closely. The fast wind quickly
overtakes the AGB envelope, drives a bow~shock on it, and a hot
bubble (10$^{7-8}$\,K) forms between the envelope and the working surface
of the fast wind. Gas is then pushed supersonically onto the
envelope, producing a compressed, spherical and efficiently-cooling
shell expanding at $\sim\,$20\,km\,s$^{-1}$.  In Simulation~3 (bottom
row), the jet forms a central bipolar cavity which is then blow
form within by the isotropic fast wind. A hot bubble forms in the
swept~up region, bound by a compressed shell that quickly adopts
an elliptical morphology and expands at~$\sim\,$20\,km\,s$^{-1}$
with a widely constant aspect ratio of~2.  Simulation~3 shows how
bipolar young PN transform into old larger elliptical nebulae, in
agreement with observed PN morphological histories.
In Simulations~4, 5~and~6 (not shown), the toroidal AGB envelope funnels
any subsequent stellar outflow towards the pole and yields narrow-waisted bi-polar or
bi-lobed objects consistently. The long term morphologies correlate
with outflow histories. An elliptical rhombus-looking shell forms
quickly and slowly expands homologically in Simulation~7 (not shown),
where the radiative hydro evolution occurs as in Sim.~2, but for
the differences in the shell shape.
In complementary simulations: gas was allowed to expands passive
between the jet and the fast wind episodes; Sim.~3 was allowed to expand for
longer (up to 26500\,yr); gas temperature was suddenly raised to 10000\,K
everywhere, to crudely simulate the effects of photoionization from the central
star. We found mild results in these experiments relative to the ones
in Table~1.

%%%%%%%%%%%%%%%%%%%%%%%%%%%%%%%%%%%%%%%%%%%%%%
\begin{figure}[ht]
  \centering
    \includegraphics[width=.20\textwidth,bb=.30in 1.5in 8in 9.5in,clip=]{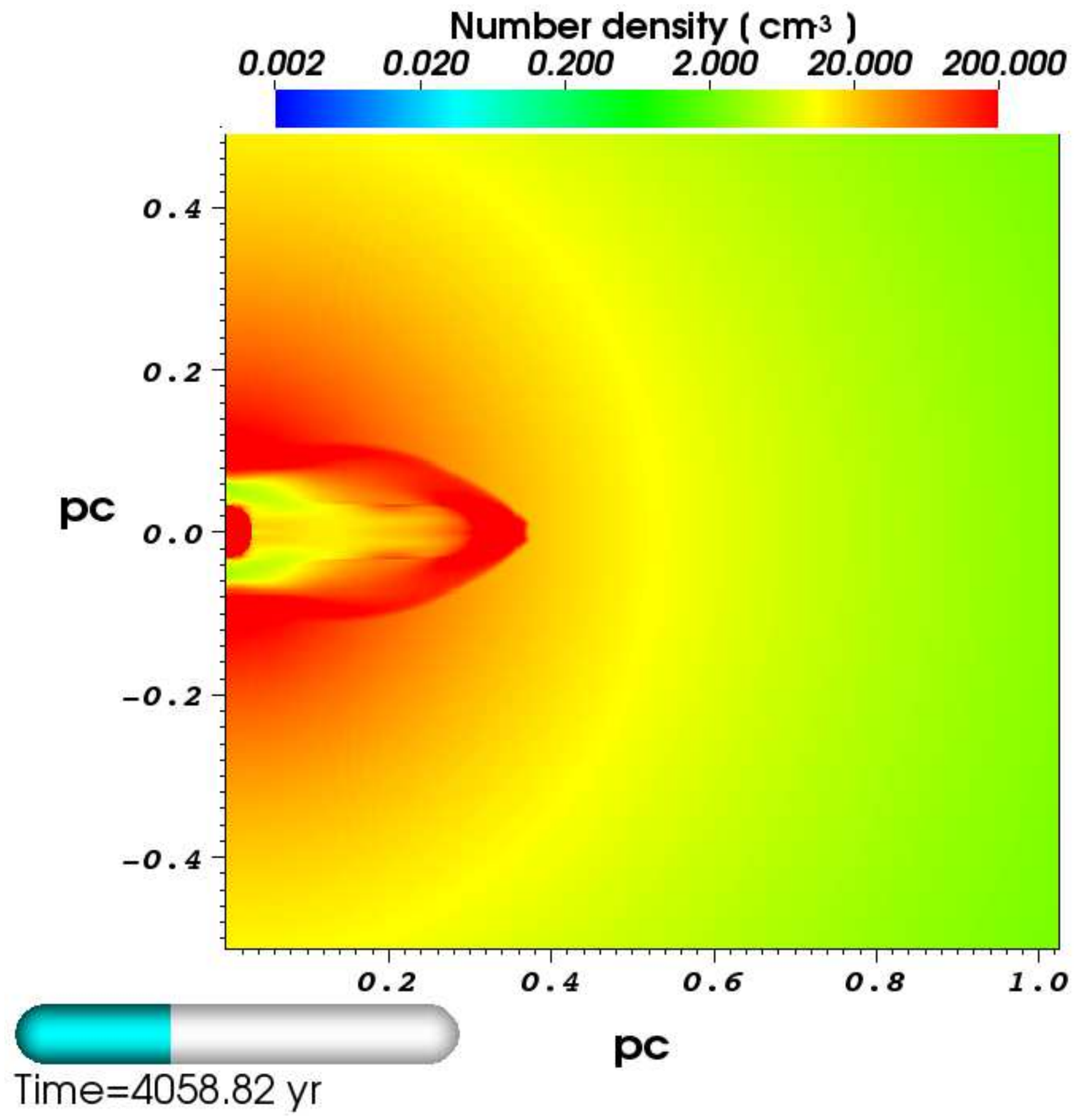}
    \includegraphics[width=.20\textwidth,bb=.30in 1.5in 8in 9.5in,clip=]{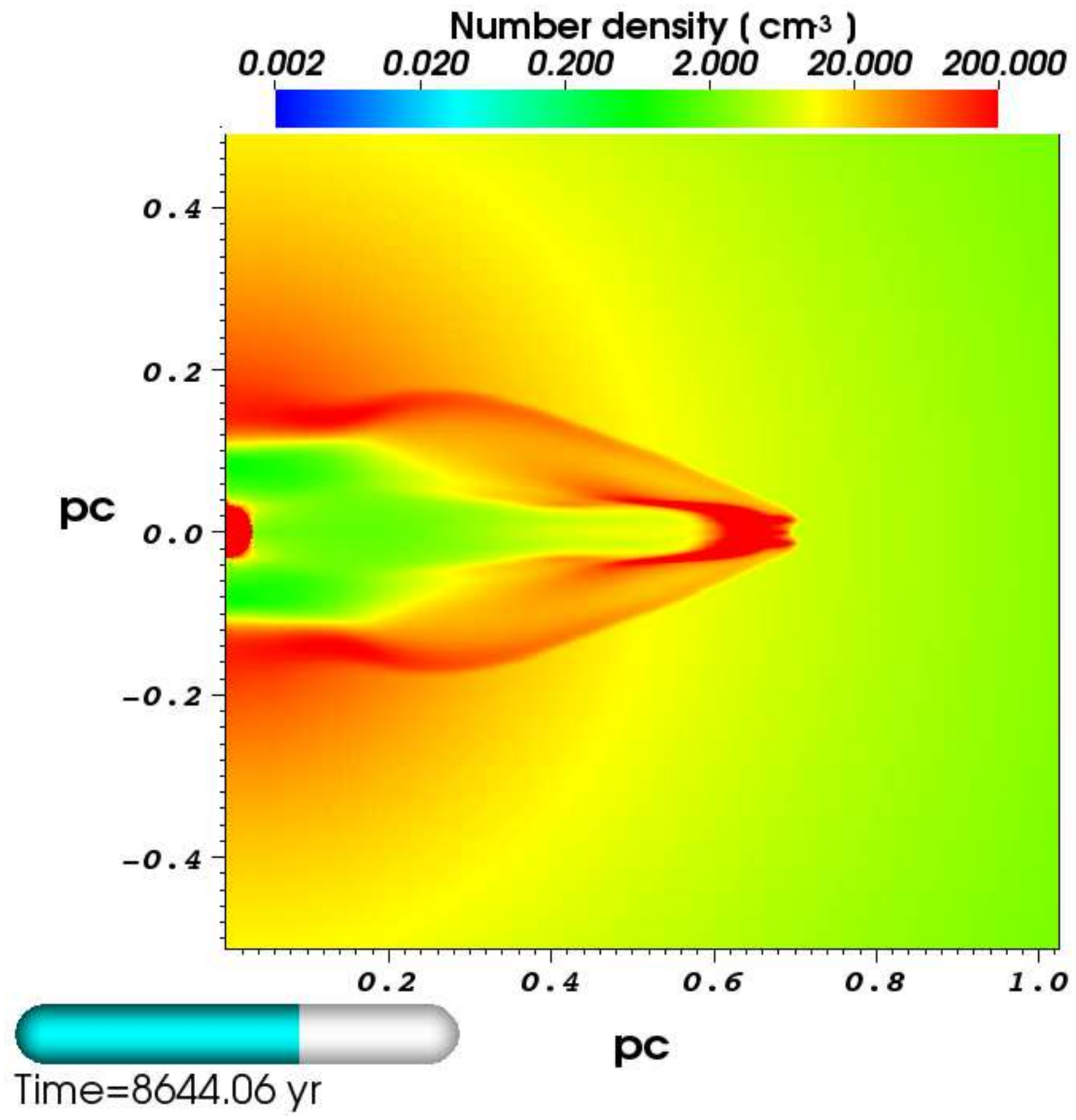}
    \includegraphics[width=.20\textwidth,bb=.30in 1.5in 8in 9.5in,clip=]{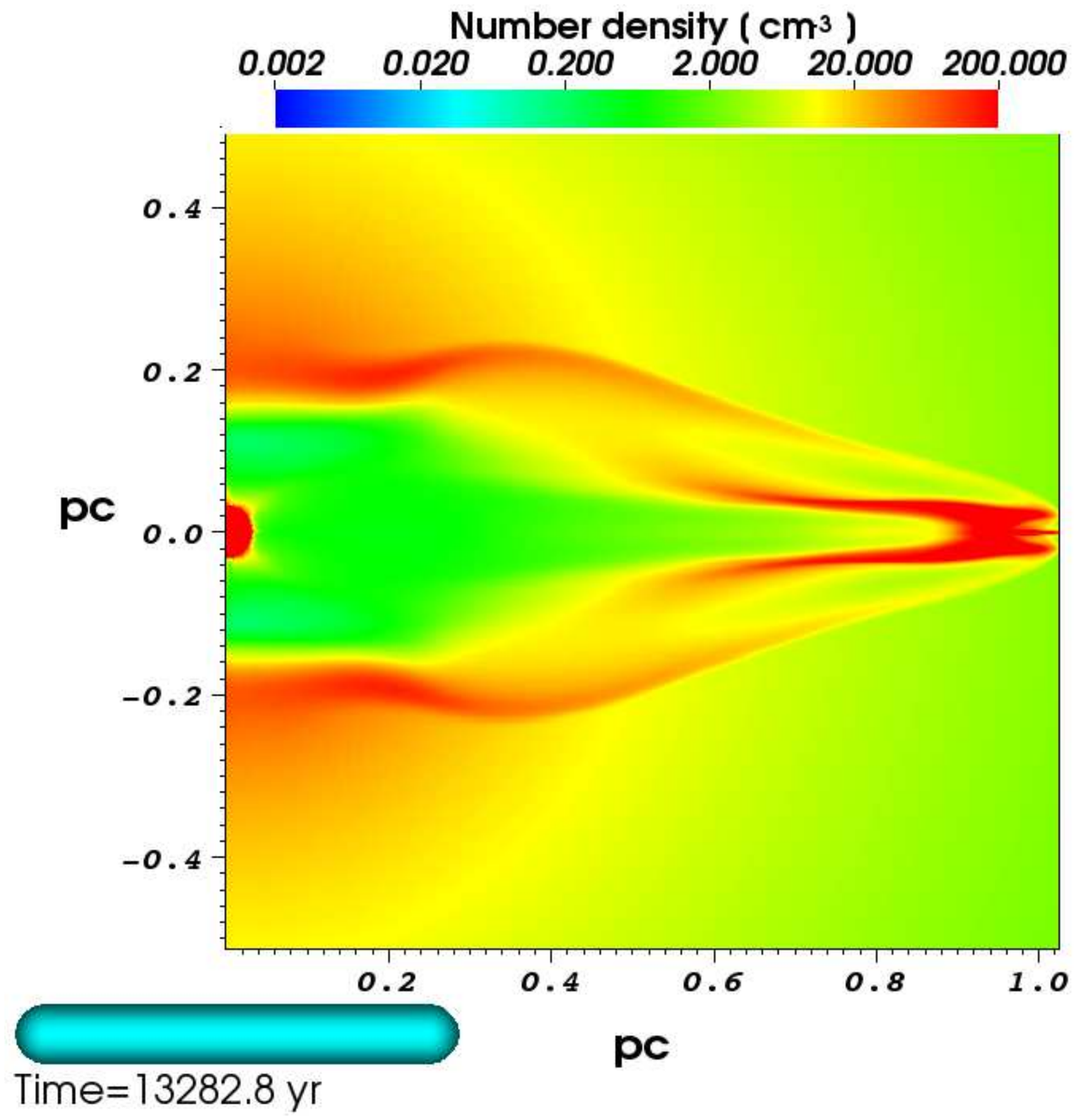}\\
    {\small Simulation~1: AGB wind $\rightarrow$ jet} \\
    \vskip0.25cm
    \includegraphics[width=.20\textwidth,bb=.30in 1.5in 8in 9.5in,clip=]{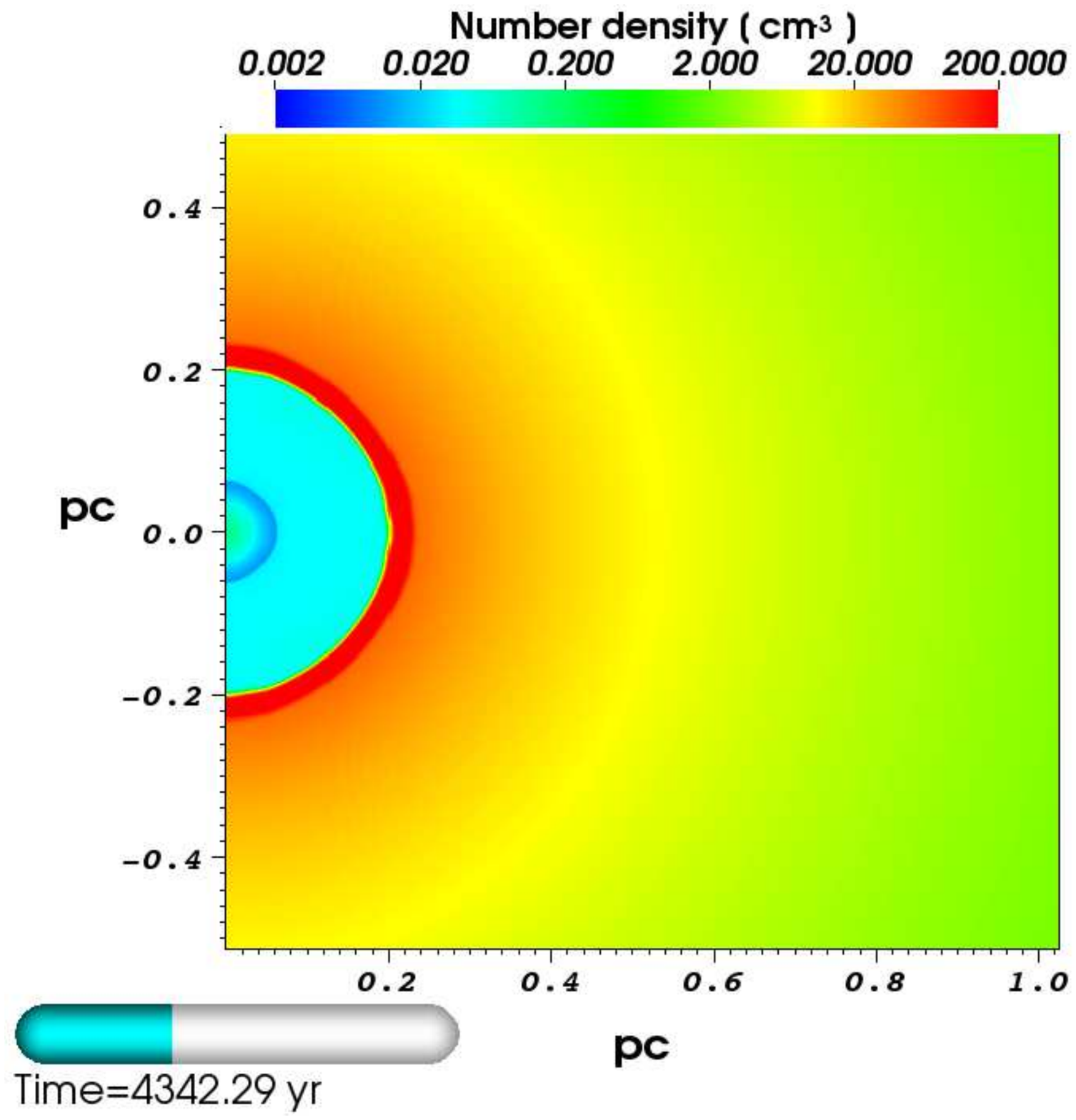}
    \includegraphics[width=.20\textwidth,bb=.30in 1.5in 8in 9.5in,clip=]{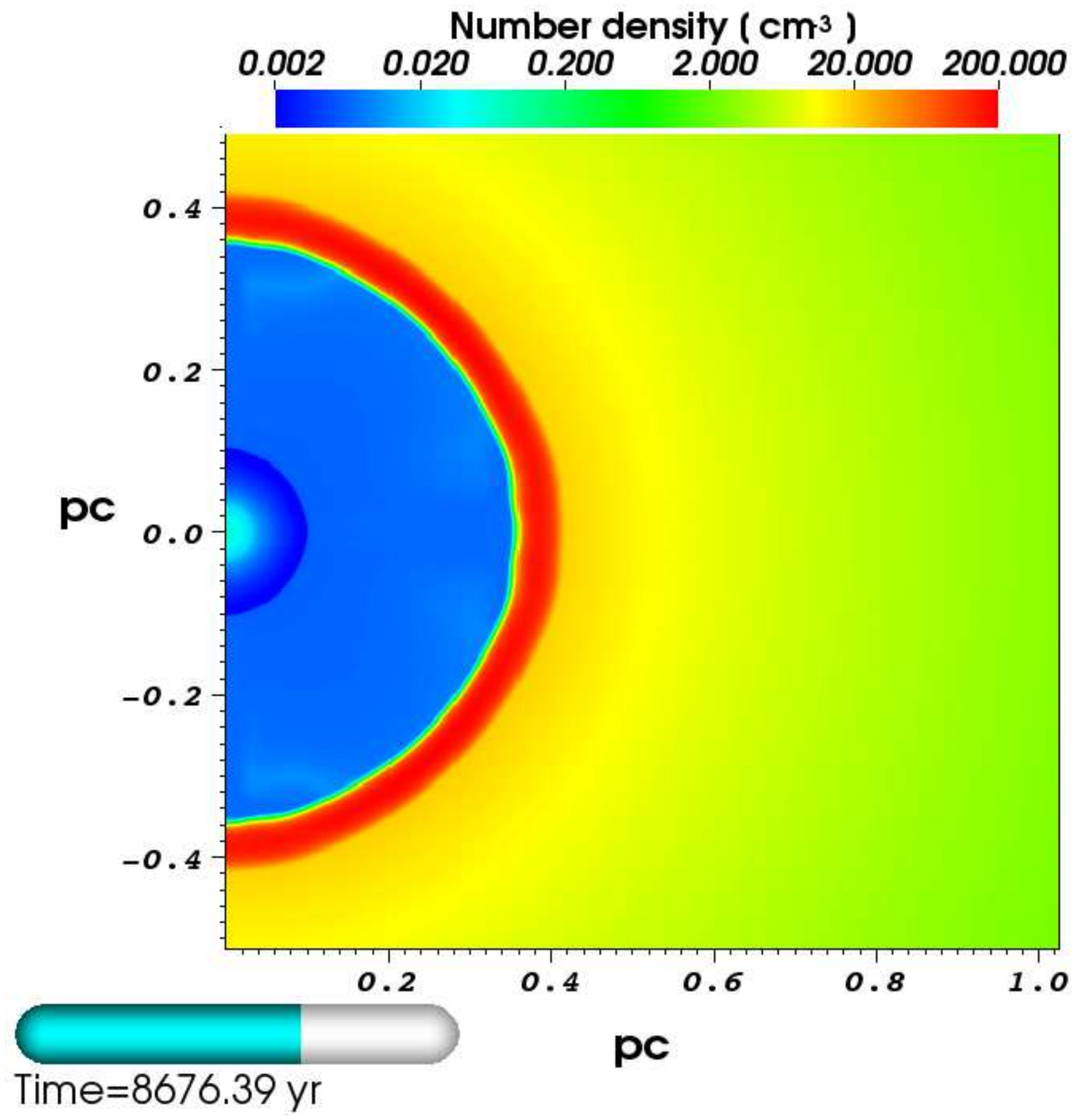}
    \includegraphics[width=.20\textwidth,bb=.30in 1.5in 8in 9.5in,clip=]{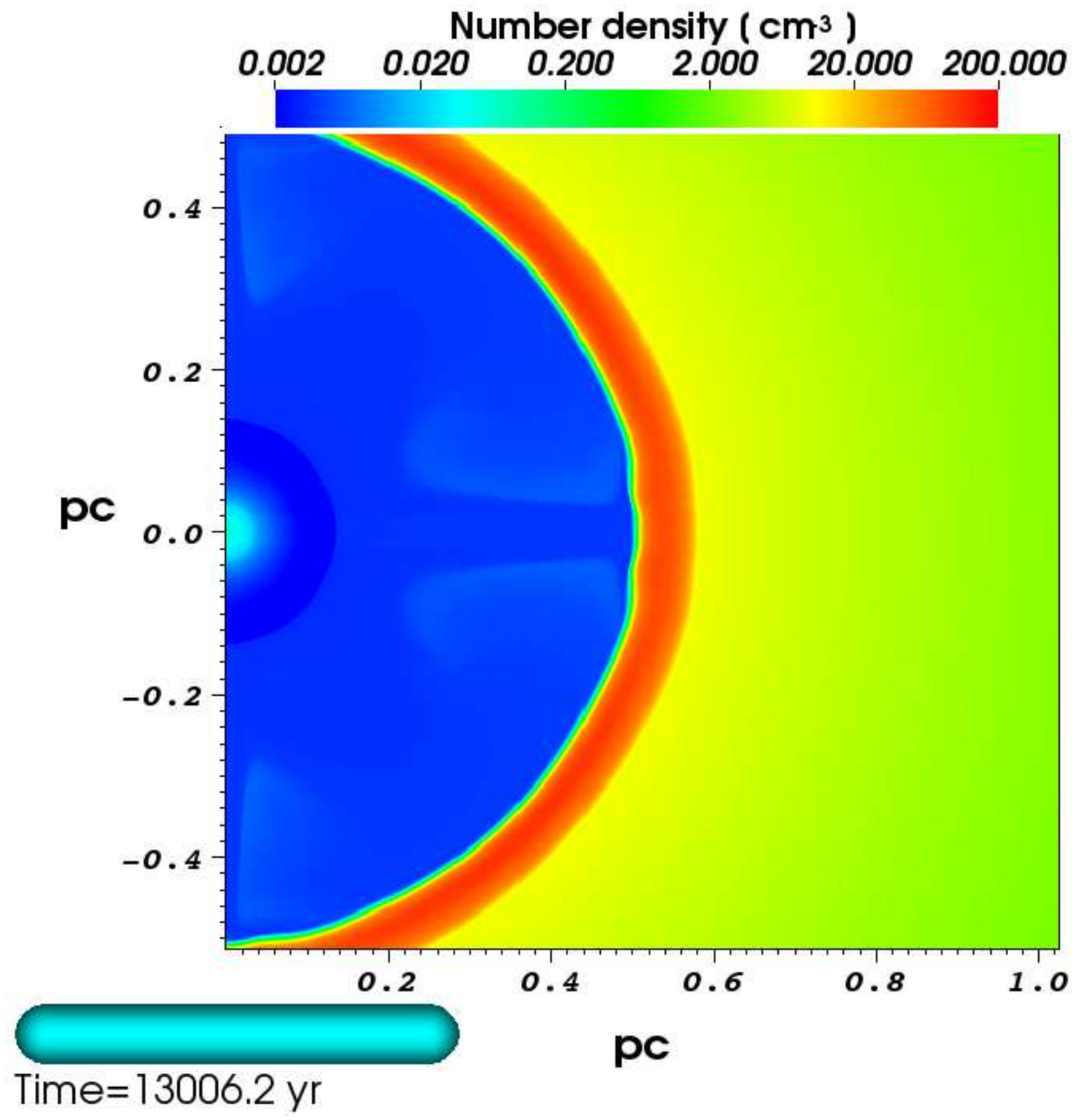}\\
    {\small Simulation~2: AGB wind $\rightarrow$ fast wind} \\
    \vskip0.25cm
    \includegraphics[width=.20\textwidth,bb=.30in 1.5in 8in 9.5in,clip=]{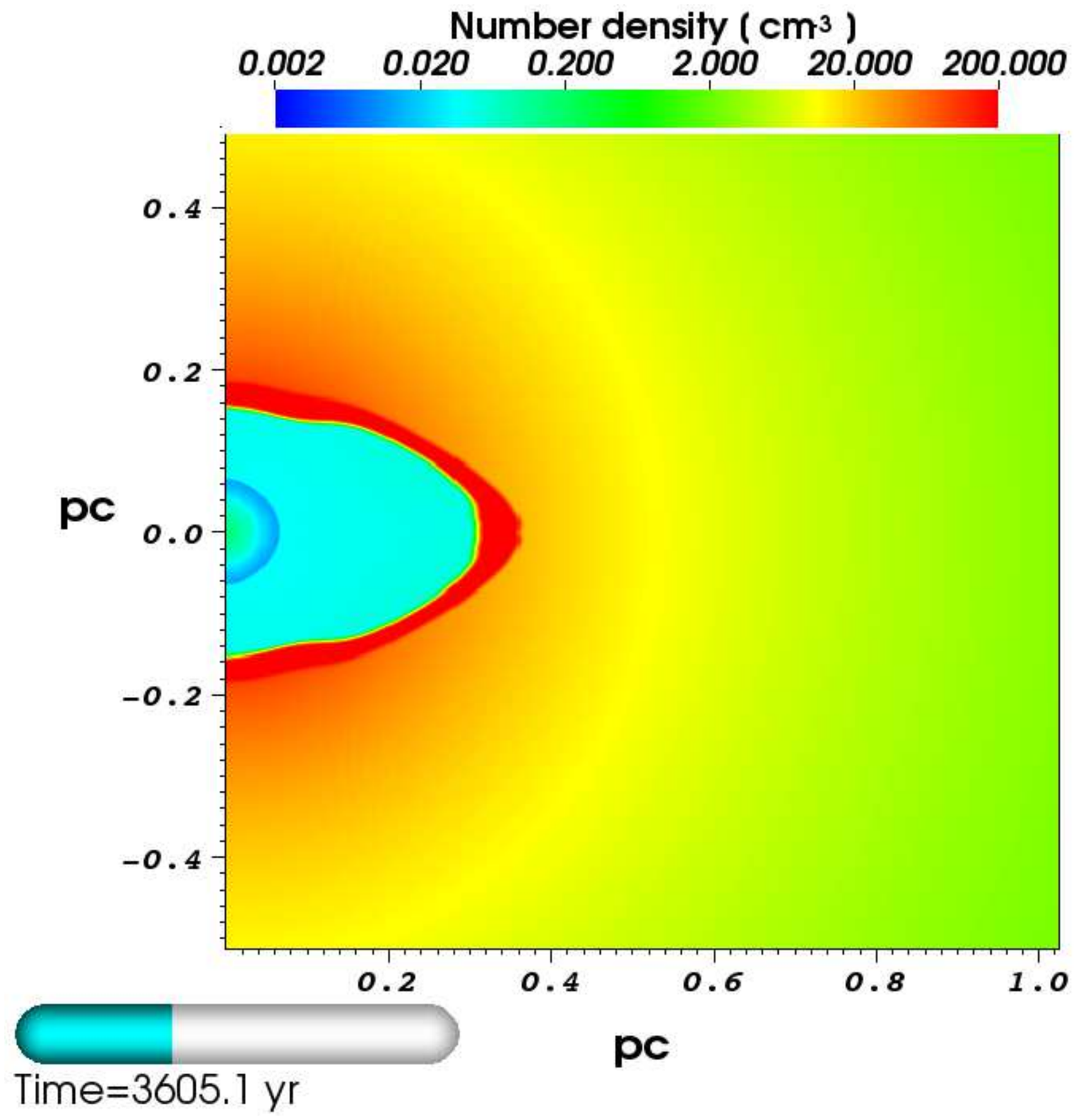}
    \includegraphics[width=.20\textwidth,bb=.30in 1.5in 8in 9.5in,clip=]{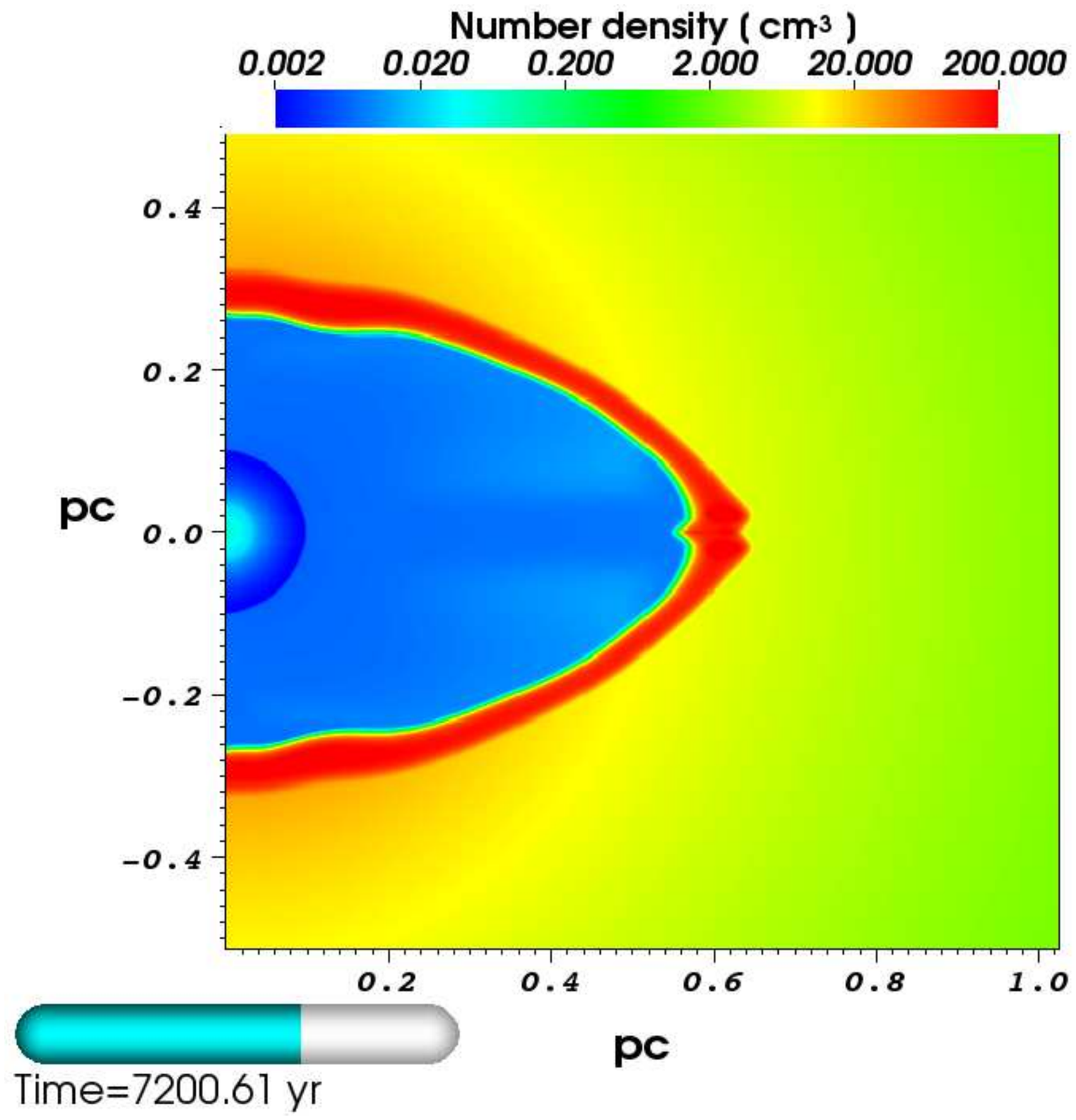}
    \includegraphics[width=.20\textwidth,bb=.30in 1.5in 8in 9.5in,clip=]{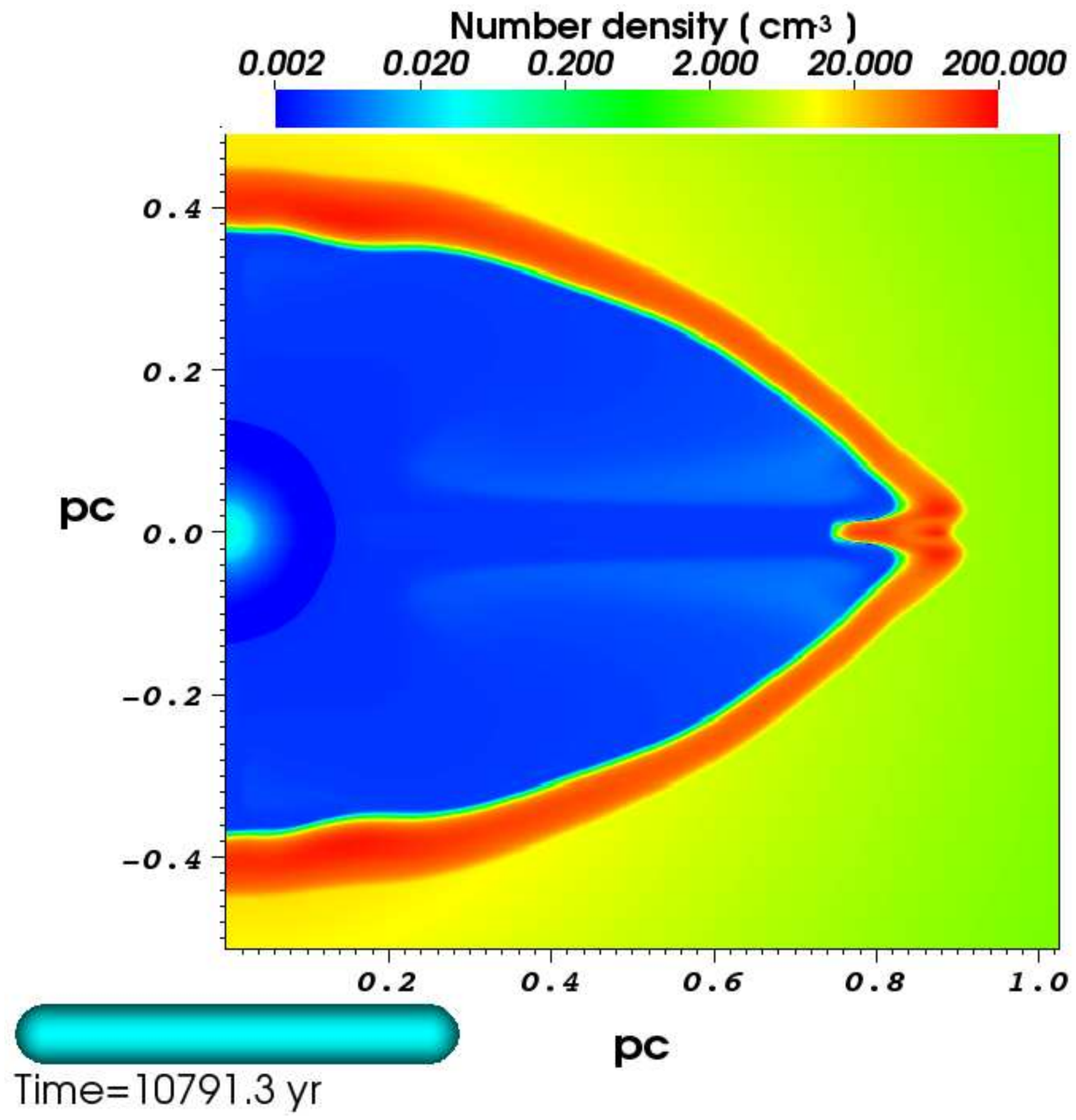}\\
    {\small Simulation~3: AGB wind $\rightarrow$ jet $\rightarrow$ fast wind}
  \caption{Evolution of the gas density in logarithmic contours.
A jet produces a narrow 
bipolar shell. An isotropic fast wind forms a spherical shell. 
A jet followed by a fast wind yields an elliptical shell. 
}
     %\vspace*{0pt}
\end{figure}
%%%%%%%%%%%%%%%%%%%%%%%%%%%%%%%%%%%%%%%%%%%%%%%

\acknowledgments The authors wish to thank the organizers of the meeting
for their work and kindness. MHE thanks Jonathan Carroll for discussions.

\bibliography{aspauthor}

\end{document}